\documentclass{SciPost}

\hypersetup{
    colorlinks,
    linkcolor={red!50!black},
    citecolor={blue!50!black},
    urlcolor={blue!80!black}
}

\usepackage[bitstream-charter]{mathdesign}
\usepackage{xspace}
\DeclareSymbolFont{usualmathcal}{OMS}{cmsy}{m}{n}
\DeclareSymbolFontAlphabet{\mathcal}{usualmathcal}

\fancypagestyle{SPstyle}{
\fancyhf{}
\lhead{\colorbox{scipostblue}{\bf \color{white} ~SciPost Physics }}
\rhead{{\bf \color{scipostdeepblue} ~Submission }}

\fancyfoot[C]{\textbf{\thepage}}
}

\newcommand{\sO}{s_{\OCAL}}
\newcommand{\sS}{s_{\SCAL}}
\newcommand{\sN}{s_{\NCAL}}
\newcommand{\tBN}{t_{\BCAL \cap \NCAL}}
\newcommand{\tAN}{t_{\ACAL \cap \NCAL}}
\newcommand{\tN}{t_{\NCAL}}
\newcommand\subfig[2]{{Fig.~\ref{#1}{#2}}}
\newcommand\subcap[1]{{(#1):}}
\newcommand{\prog}[1]{Alg.~\ref{alg:#1} (\sub{#1})}

\newcommand{\progg}[1]{Algorithm~\ref{alg:#1} (\sub{#1})}
\newcommand{\proggg}[1]{Algorithm \sub{#1}}

\newcommand{\ZRA}[2]{Z_{#1}^{\{#2\}}}
\newcommand{\VR}{variation ratio\xspace}

\newcommand{\REF}[2][]{
        \ifthenelse{\equal {#1} {}}{Ref.~\cite{#2}}{Ref.~\cite[#1]{#2}}}

\newcommand{\PROCEDURE}[1]{\textbf{procedure}\ \sub{#1}}

\newcommand{\BRACE}[1]{
    \;\;\;  \left\{\begin{array}{l}#1\end{array} \right.}
\newcommand{\IS}[2]{#1 \leftarrow #2}

\newcommand{\ENDPROCEDURE}{\text{------} \\ \vspace{-0.8cm}}
\newcommand{\IF}[1]{\textbf{if } #1 \textbf{: }}

\newcommand{\OUTPUT}[1]{\textbf{output}\ #1}

\newcommand{\INPUT}[1]{\textbf{input}\ #1}
\newcommand{\COMMENT}[1]{\text{\footnotesize (#1)}}

\newcommand{\SET}[1]{\{#1\}}

\newcommand{\sub}[1]{\texttt{#1}}
\newcommand{\eq}[1]{eq.~\eqref{#1}}

\newcommand{\eqtwo}[2]{eqs~\eqref{#1} and~\eqref{#2}}

\newcommand{\fig}[1]{Fig.~\ref{#1}}
\newcommand{\figg}[1]{Figure~\ref{#1}}
\newcommand{\quot}[1]{``#1''}
\newcommand{\tab}[1]{Table~\ref{#1}} 
 
\newcommand{\sect}[1]{Section~\ref{#1}} 
\newcommand{\app}[1]{Appendix~\ref{#1}}

\newcommand{\etc}{\textrm{etc.}}

\newcommand{\ACAL}{\mathcal{A}}  %  mathcal 
\newcommand{\BCAL}{\mathcal{B}}  %  mathcal
\newcommand{\MCAL}{\mathcal{M}}  %  mathcal
\newcommand{\NCAL}{\mathcal{N}}  %  mathcal
\newcommand{\OCAL}{\mathcal{O}}  %  mathcal
\newcommand{\SCAL}{\mathcal{S}}  %  mathcal
\newcommand{\expb}[1]{\exp \glb #1 \grb} % low exponential with groupings b
\newcommand{\expc}[1]{\exp \glc #1 \grc} % low exponential with groupings c
\newcommand{\ran}{\texttt{ran}}
\newcommand{\ranb}[2][]{\ran_{#1} \! \glb #2 \grb}  % ran-brace,  with - nothing
\newcommand{\glb}{\left(}  % ' group left b' 
\newcommand{\grb}{\right)}  % ' group right b' 
\newcommand{\glc}{\left[}  % ' group left c'
\newcommand{\grc}{\right]}  % ' group right c' 
\newcommand{\gld}{\left\{}  % ' group left d' 
\newcommand{\grd}{\right\}}  % ' group right d' 
\newcommand{\gle}{\left|}  % ' group left d' 
\newcommand{\gre}{\right|}  % ' group right d' 

\newcommand{\const}{\text{const}}

\newcommand{\PLUSPLUS}{+ \dots +}

\newcommand{\TO}{,\ldots,}

\newcommand{\mean}[1]{\left\langle #1 \right\rangle}
\newcommand{\half}{\frac{1}{2}}

\newcommand{\fpn}[2]{\ensuremath{#1 \! \times \! 10^{#2}}}
\newcommand{\fracd}[2]{\dfrac{#1}{#2}}

\newcommand{\draftfigure}[4][\linewidth]{\begin{figure}[htbp]
   \begin{center}
      \includegraphics[width=#1]{#2}
   \end{center}
   \caption{#3}
   \label{#4}
\end{figure}}

\usepackage{float}
\usepackage{comment}
\begin{document}
\newfloat{algorithm}{ht}{loa}
\floatname{algorithm}{Algorithm }
\pagestyle{SPstyle}

\begin{center}{\Large \textbf{\color{scipostdeepblue}{
%%%%%%%%%% TODO: Write your article's title here
Thermodynamic integration, fermion sign problem, and real-space
renormalization\\
}}}\end{center}

\begin{center}\textbf{
% Use (full) first name (+ middle name initials) + surname format.
% Separate subsequent authors by a comma, omit comma and use "and" for the last author.
% Mark the corresponding author(s) with a superscript symbol in this order
% \star, \dagger, \ddagger, \circ, \S, \P, \parallel, ...
Koka Sathwik\textsuperscript{1$\dagger$} and
Werner Krauth\textsuperscript{1, 2$\star$}
}\end{center}

\begin{center}
{\bf 1} Laboratoire de Physique de l’Ecole normale supérieure, ENS,
Université PSL, CNRS, Sorbonne Université, Université Paris-Diderot, Sorbonne
Paris Cité, Paris, France
\\
{\bf 2} Rudolf Peierls Centre for Theoretical Physics, Clarendon
    Laboratory, Oxford OX1 3PU, UK
%%%%%%%%%% END TODO: AFFILIATIONS
% Provide email address of corresponding author(s)
\\[\baselineskip]
$\dagger$ \href{mailto:koka.sathwik@ens.psl.eu}{\small koka.sathwik@ens.psl.eu}
$\star$ \href{mailto:werner.krauth@ens.fr}{\small werner.krauth@ens.fr}
\end{center}

\section*{\color{scipostdeepblue}{Abstract}}
\boldmath\textbf{%
%%%%%%%%%% TODO: ABSTRACTth
We reconsider real-space renormalization for the two-dimensional Ising model,
following the path traced out by Wilson in Sect. VI of his 1975 Reviews of
Modern Physics~\cite{Wilson1975}. In that reference, Wilson considerably
extended the Kadanoff decimation procedure towards a possibly rigorous
construction of a real-space scale-invariant hamiltonian.
Wilson's construction has, to the best of our knowledge, never
been fully understood and thus neither reproduced nor generalized. In the
present work, we use Monte Carlo sampling in combination with thermodynamic
integration in order to retrace Wilson's computation for a real-space
renormalization with a number of terms in the hamiltonian. We
elaborate on the connection of real-space renormalization with the fermion sign
problem and discuss to which extent our Monte Carlo procedure actually
implements Wilson's program from half a century ago.
}

\vspace{\baselineskip}

%%%%%%%%%% BLOCK: Copyright information
% This block will be filled during the proof stage, and finilized just before publication.
% It exists here only as a placeholder, and should not be modified by authors.
\noindent\textcolor{white!90!black}{%
\fbox{\parbox{0.975\linewidth}{%
\textcolor{white!40!black}{\begin{tabular}{lr}%
  \begin{minipage}{0.6\textwidth}%
    {\small Copyright attribution to authors. \newline
    This work is a submission to SciPost Physics. \newline
    License information to appear upon publication. \newline
    Publication information to appear upon publication.}
  \end{minipage} & \begin{minipage}{0.4\textwidth}
    {\small Received Date \newline Accepted Date \newline Published Date}%
  \end{minipage}
\end{tabular}}
}}
}
%%%%%%%%%% BLOCK: Copyright information

% For convenience during refereeing we turn on line numbers:
%\linenumbers

\vspace{10pt}
\noindent\rule{\textwidth}{1pt}
\tableofcontents
\noindent\rule{\textwidth}{1pt}
\vspace{10pt}

% Write your article contents here, starting from first \section.
% An example structure is given below.

\section{Introduction}
\label{sec:Intro}
The renormalization group~\cite{Wilson1975}, a cornerstone of theoretical
physics, originates in the concept of universality in statistical mechanics. It
implements that the physics on large length scales, at a critical point, is
described through  a few models that differ qualitatively, for example through
the symmetries of their hamiltonians. A renormalization procedure computes
properties of the system from the constraint of having to arrive at such a
distinct model in the limit. A prominent variant of the theory is real-space
renormalization. It was pioneered  by Kadanoff~\cite{Kadanoff1975} as a
decimation procedure in which half of the  spins are eliminated (integrated
out), and the remaining ones identified from one iteration of the
renormalization procedure to the next one. As analyzed by Wilson, this
renormalization consists in computing the partition function at the critical
point, where the correlation length becomes infinite, and where spin--spin
correlations decay as power laws. At this critical point, the fixpoint
hamiltonian of the system is described by universal coupling constants. Despite
its conceptual appeal as a direct application of scale invariance, real-space
renormalization has mainly been presented in a pedagogical
context~\cite{Maris1978}, with only a few quantitative
applications~\cite{BellWilson1975,Griffiths1978,Kennedy1993,Kennedy2010,
vanEnter1993,Haller1996} . Nevertheless, Wilson has shown \REF[Section
VI]{Wilson1975} that real-space renormalization can avoid the collapse of
spin--spin correlations by decimating (as before) half of the spins but without
identifying the remaining spins over the iterations of the renormalization. This
explicitly constructs the scale-invariant hamiltonian and allows for the
computation of critical exponents. Wilson's computation for the Ising model has
been notoriously difficult to reproduce~\cite{Rychkov2023}, and has not been
generalized. The present work rephrases Wilson's computation in the language of
modern computational physics and reproduces some of its results by unbiased
Monte Carlo sampling methods that involve thermodynamic integration~\cite{SMAC}
and that treat an emergent fermionic sign problem~\cite{Blankenbecler1981}. This
work is self-contained (see \app{app:WilsonVITwoParameter}). Our algorithms
merely aim at a  proof of concept. They are implemented as Python scripts and
are openly available (see \app{app:AccessComputer}).

\section{Decimation---real-space renormalization}

Real-space renormalizion consists, in a nutshell, in setting up a very general
hamiltonian for the interaction of Ising spins and then in integrating out a
certain fraction of the spins and in writing the result as the same hamiltonian.
In real space,  this hamiltonian  must have many more terms beyond the
usual nearest-neighbor interaction of the Ising model which, by itself, is not
invariant under scale transformations.

We use factors to generalize the pairs of nearest-neighbor spins in the usual
Ising model. Each factor $M$ is defined through a set $I_M$ of site indices
on which it couples the spins.
The index set $I_M$
implies a certain type $T_M$ (two neighboring sites, two next-nearest
neighbors, four sites around a fundamental plaquette, \etc.). For clarity
and ease of book-keeping, we write each factor as $M := (I_M, T_M)$, although
the type is implied by the index set. Index sets $I_M$ and
$I_{M'}$ that are related by rotations, reflections and
translations satisfy $T_{M}= T_{M'}$. An interaction $K_T$ is associated with
each factor type $T$, so that the energy of  a configuration of spins
$\sS$ is:
\begin{equation}
H(\sS) =  \sum_{M = (I_M, T_M) \in \MCAL} K_{T_M} \prod_{j \in
I_M} s_j.
\label{equ:Hamiltonian}
\end{equation}
The number of possible factors is huge, potentially as big as the powerset (the
set of subsets) of all indices. \REF[Table III]{Wilson1975} provides $14$
dominant factor types $T \in \SET{1 \TO 14}$ (see \fig{fig:FactorTypes}).
We consider two cases, one where the set $\MCAL$ of factors with non-zero
weight have $T\in \SET{1,2}$ (nearest and next-nearest neighbors), and the
other where all $14$ factor types in \fig{fig:FactorTypes} are taken into
consideration.

\draftfigure[10cm]{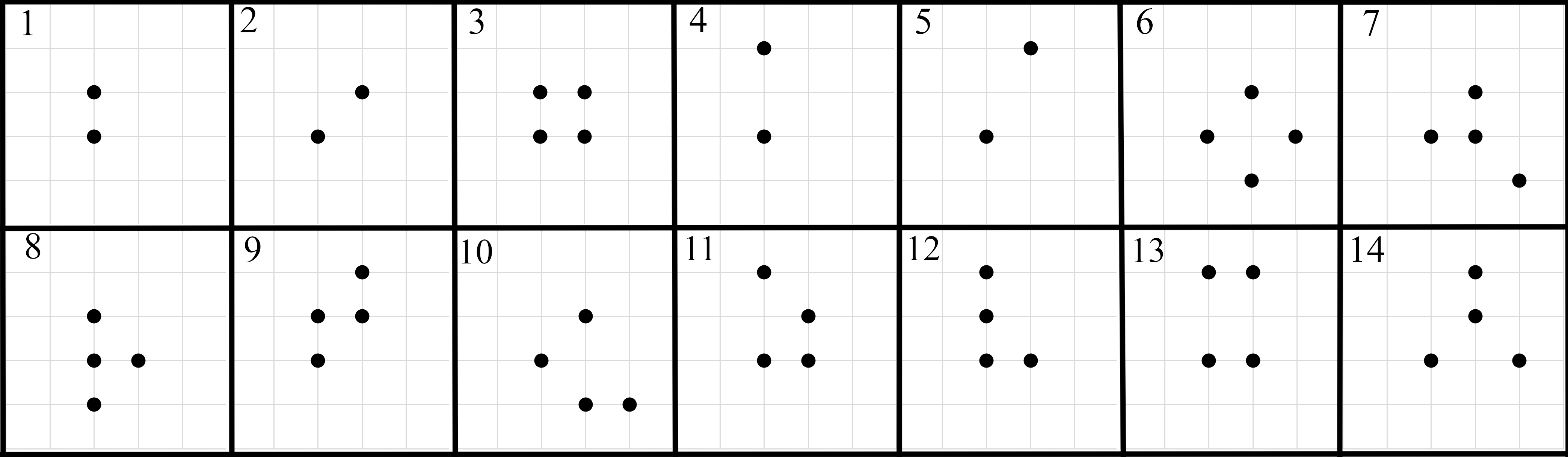}{ The $14$ dominant factor types
$T \in \SET{1 \TO 14}$ identified in \REF[Table III]{Wilson1975}.
In the Wilson patch of old sites (set $\SCAL$ of
\eq{equ:SDefinition}), they appear in $9447$ factors. In the patch of new
sites (set $\NCAL$ of \eq{equ:NDefinition}), they appear in $4795$
factors.}{fig:FactorTypes}

\subsection{Wilson patch}
\label{sec:WilsonPatch}

\figg{fig:Wilsonfig8}, reproduced from \REF[Fig. 8]{Wilson1975}, defines the
\emph{Wilson patch} to which we restrict our attention throughout this work. It
consists in a finite two-dimensional square lattice with, at iteration $i$, a
set $\SCAL$ of  $221$ sites, of which $17$ sites are located inside the
\emph{old central region} $\BCAL$, with the rest in the \emph{old periphery}
$\ACAL$.
In the next iteration, $i+1$, a set $\NCAL$ of $121$ \emph{new} sites remains,
with
$11$ in the \emph{new central region} $\BCAL \cap \NCAL$. Our numbering scheme
assigns the same indices for $\SCAL$ and for $\NCAL$:
\begin{align}
\SCAL &= \SET{1 \TO 221}             & |\SCAL| &= 221 & \text{old
sites} \label{equ:SDefinition},\\
\OCAL &= \SET{12, 13 \TO 21, \dots } & |\OCAL| &= 100 &  \text{\quot{old-only}
sites},
\\
\NCAL &= \SCAL \setminus \OCAL & |\NCAL| &= 121 & \text{new sites},
\label{equ:NDefinition}
\\
\BCAL &= \SET{ 89, 90 \TO 134} & |\BCAL| &= 17& \text{old central sites},\\
\ACAL &= \SCAL \setminus \BCAL & |\ACAL| &= 204 & \text{old \quot{peripheral}
sites},\\
\BCAL \cap \NCAL &= \SET{ 89, 90 \TO 133, 134},
&|\BCAL \cap \NCAL| &= 11& \text{new central sites}.
\label{equ:BasicDefinitions}
\end{align}
Following the notations in \REF{Wilson1975}, we refer to the \quot{old}
spins as
\quot{$s \in \SET{-1,1}$} and to the \quot{new} spins as \quot{$t \in
\SET{-1, 1}$}, writing for example $\sS
= \SET{s_i: i \in \SCAL}$
for the set of old spins and $\tN = \SET{t_i: i \in \NCAL}$ for the
set of new spins.
Likewise, we write $\tBN = \SET{t_i: i \in \BCAL \cap \NCAL}$ for the
new central spins, \etc. The $2^{11}$ spin configurations in the new
central region $\BCAL \cap \NCAL$ are
numbered using the Gray code (see \app{app:Details}).

\draftfigure[6cm]{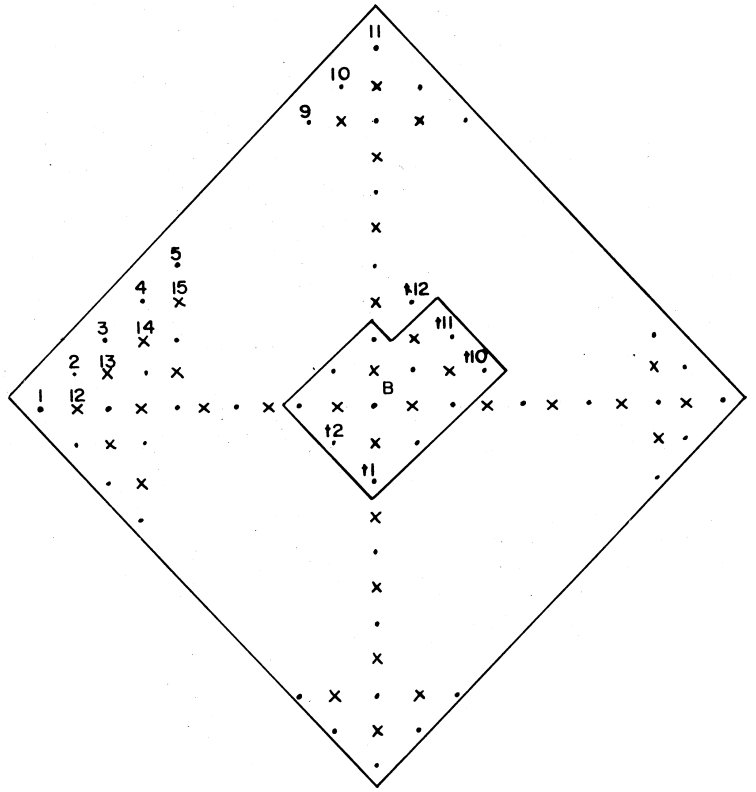}{Wilson patch from
\REF{Wilson1975}. Naively, the old-only spins---the \quot{$\times$} in the
picture---are to be decimated (summed over) (see
\sect{sec:OriginalDecimation}),
yielding a partition function and at the same time the Boltzmann
weight of the new spins---the \quot{$\cdot$} in the picture. At the fixpoint,
the partition function over the old spins agrees with the Boltzmann weight of
the new spins. The naive procedure has a flaw, which was corrected by Kadanoff
(see \sect{sec:ModifiedDecimation}).
}{fig:Wilsonfig8}

\subsection{Kadanoff's original decimation, flaw}
\label{sec:OriginalDecimation}
Kadanoff's original decimation consists in integrating out (that is, summing
over) the old-only spins $\sO$ for a fixed configuration of new spins:
\begin{equation}
Z (\underbrace{\tBN | \tAN}_{
\text{fix all new spins}\ \tN}
) = \sum
_{\sO| \sN = \tN}
\expc{ H(\sS)}.
\label{equ:WilsonPartitionEq5}
\end{equation}
On the Wilson patch of \fig{fig:Wilsonfig8}, the sum on the right-hand side of
the above equation involves $|\sO| = 100$ \quot{old-only} spins, and thus runs
over $2^{100}$ terms. All the new  spins $t$ are then fixed, so that the above
partitition function corresponds to a single configuration of $t$-spins. At the
fixpoint, the latter is the exponential of  the hamiltonian for the $t$ spins:
\begin{equation}
Z (\tBN | \tAN ) \stackrel{!}{=} \expc{H(\tBN | \tAN )}.
\end{equation}
In \REF{Wilson1975}, Wilson fixes the outer spins  as $\tAN = \SET{1 \TO 1}$.
Varying the $11$ new spins $\tBN$ then leaves $2^{11}$ equations to be
satisfied, which allows one to search for a fixpoint.

The decimation just described has a flaw that arises from the identification
of old spins that survive the decimation with the new spins, in other words from
the condition $\sN = \tN$ on the right-hand side of \eq{equ:WilsonPartitionEq5}.
The identification impacts the scaling of spin--spin
correlation functions at the critical point.  Indeed, as first understood by
Kadanoff (cited in \REF{Wilson1975}), certain lattice points belong to many
iterations of the renormalization procedure and under the original
decimation, the spins and the spin--spin correlations are identical across
iterations. This implies that the spin--spin correlations must be
independent of distance and, in fact, zero. At the critical point, where one
knows that the decay of correlations is algebraic, this is inconsistent.

\subsection{Modified decimation, coupling parameter $\rho$}
\label{sec:ModifiedDecimation}

To repair the flaw, one considers, following Kadanoff, a decimation procedure
that does not identify the surviving old spins with the new spins. The old-only
spins are still integrated out. New and old spins on $\NCAL$, that is, $\sN$ and
$\tN$, are coupled through the following ansatz (see \fig{fig:CouplingSAndT}):
\begin{equation}
Z_\rho(\tBN | \tAN) =
\sum_{\sS} \gld \expc{H(\sS)} \prod_{k \in \NCAL} (1+ \rho t_k
s_k)\grd.
\label{equ:ModifiedDecimation}
\end{equation}
Here, the coupling parameter $\rho$ is larger than one. As $t_k,
s_k \in
\SET{-1,1}$, the final term on the right-hand side of
\eq{equ:ModifiedDecimation} can thus be positive or negative, and $Z_\rho$
resembles a fermion partition function.

\draftfigure[10cm]{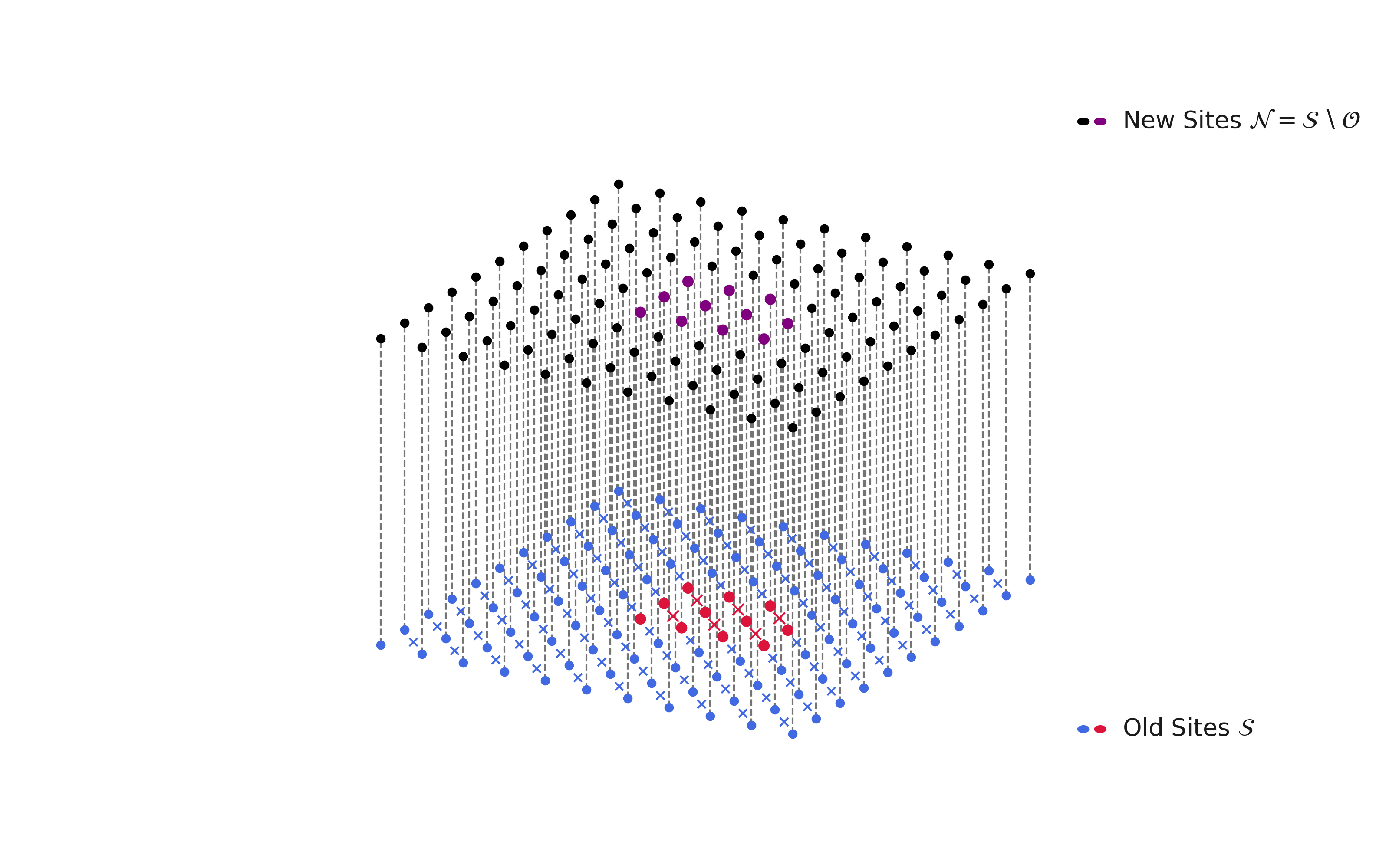}{Wilson patch with the modified decimation. All
$s$ spins (lower-level crosses and dots) are summed over in order to compute
the effective interaction between $t$ spins (upper-level dots). Spins $s_k$ and
$t_k$, for $k \in \NCAL$ (dots on the same sites on both levels) are coupled by
$1 + \rho t_k s_k$ (vertical lines), but not identified (see
\eq{equ:ModifiedDecimation}). }{fig:CouplingSAndT}

\section{Markov-chain Monte Carlo procedure}
\label{sec:ModifiedMonteCarloProcedure}

Wilson, in \REF{Wilson1975}, approximately evaluated $Z_\rho$ in
\eq{equ:ModifiedDecimation}, using a procedure that was not fully explained. We
compute this partition function exactly using Monte Carlo methods. In the
present section, we discuss the three main ingredients of this computation. The
first is thermodynamic integration (\sect{sec:ThermodynamicIntegration}) which
allows one to compute a partition function, rather than the expectation of an
observable. The second ingredient is the treatment of the fermion sign problem
(\sect{sec:FermionSign}) that arises from the presence, for $\rho > 1$,  of
negative contributions to the partition function. On the Wilson patch, the
fermion sign problem is not generally severe, and we can evaluate $Z(\tBN |
\tAN)$ because negative and positive contributions do not cancel and the average
sign remains non-zero. Finally, we discuss correlated sampling
(\sect{sec:CorrelatedSampling}), which is crucial because we ultimately have to
solve a fixpoint equation, in other words a minimization problem. At the
minimum, to be determined, we must state that the solution is \quot{better} than
at all nearby values in a high-dimensional space. Correlated sampling allows one
to make that statement with vanishing statistical errors. We also point out
other potential uses of correlated sampling.

\subsection{Thermodynamic integration}
\label{sec:ThermodynamicIntegration}
The fixpoint equation requires the computation of the partition function
$Z_\rho$, in other words the computation of a free energy. This requires
thermodynamic integration (see \cite[Sect. 1.4.2]{SMAC} for an
introduction). In essence, given two unnormalized weight functions $\pi_x$ and
$\pi'_x$ in a given sample space $\Omega$, neither the partition function $Z =
\sum_{x \in \Omega} \pi_x$ nor the partition function $Z' = \sum_{x\in \Omega}
\pi'_x$ can be sampled directly without visiting all of $\Omega$. However, the
ratio of the two partition functions is given by
\begin{equation}
 \frac{Z'}{Z}  = \frac{\sum_x \pi'_x}{\sum_x \pi_x} = \frac{\sum_x \pi_x
\glb \pi'_x / \pi_x \grb}{\sum_x \pi_x} = \mean{\frac{\pi'_x}{\pi_x}}_{\pi},
\end{equation}
that is, by an average of the observable $\pi' / \pi$ over the
distribution
$\pi$, which can be obtained through Monte-Carlo sampling. In practice, to
compute $Z'/Z$ with high precision, the fluctuations of $\pi'(x) / \pi(x) $
and, thus, the
difference between $\pi$ and  $\pi'$ must be small.

In our context, we consider the auxiliary partition function with an additional
parameter $\alpha$ allowing for small differences between nearby parameters:
\begin{equation}
\ZRA{\rho}{\alpha}(\tN) =
\sum_{\sS} \gld \expc{\alpha H(\sS)} \prod_{k \in \NCAL} (1+ \rho t_k
s_k)\grd, \quad 0\le \alpha \le 1.
\label{equ:ModifiedDecimationAlpha}
\end{equation}
This auxiliary partition function takes on the values
\begin{equation}
\ZRA{\rho}{\alpha}(\tN) =
\begin{cases}
2^{|\SCAL|}& \text{for $\alpha = 0$} \\
Z_\rho(\tN) & \text{for $\alpha=1$ (see \eq{equ:ModifiedDecimation})}.
\end{cases}
\label{equ:ModifiedDecimationAlphaSpecial}
\end{equation}
The case $\alpha=0$ corresponds, in a free-energy computation, to the
high-temperature or zero-density limit of a physical system, where the free
energy is known analytically. The parameter $\alpha$ is thus a fictitious
inverse temperature.

To compute $Z_\rho$,  we consider a sequence of $n$  values of
$\alpha$:
$1 = \alpha_1 > \alpha_2 > \cdots > \alpha_n = 0 $
and, likewise, a sequence of $n$ values of $\rho$:
$\rho = \rho_1 \simeq \rho_2 \simeq \cdots \simeq\rho_n$
and write the telescopic product as
\begin{equation}
\ZRA{\rho}{\alpha=1}(\tN) =
 \frac{\ZRA{\rho_1}{\alpha_1}} {\ZRA{\rho_2}{\alpha_2}}
 \frac{\ZRA{\rho_2}{\alpha_2}} {\ZRA{\rho_3}{\alpha_3}}
 \dots
 \frac{\ZRA{\rho_{n-1}}{\alpha_{n-1}}} {\ZRA{\rho_n}{\alpha_n}}
 \ZRA{\rho_n = 0}{\alpha_n},
\label{equ:TelescopicProduct}
\end{equation}
where, on the right-hand side, we have omitted the dependence on $\tN$.
Each of the ratios in this expression can be sampled in principle, but not in
practice, because of the presence of positive and negative terms in the
partition function.

\subsection{Fermion sign problem}
\label{sec:FermionSign}
For $\rho> 1$, the sum in \eq{equ:ModifiedDecimationAlpha} has positive and
negative contributions, and the individual terms can thus not be interpreted as
probabilities. Following a classic reference by Blankenbecler et
al.~\cite{Blankenbecler1981}, we therefore replace the term $\Pi_\rho: =\prod_{k
\in \NCAL} (1+ \rho t_k s_k)$ by its absolute value. Dividing by and
multiplying with $|\Pi_\rho|$, any of the ratios in \eq{equ:TelescopicProduct}
can be written as:
\begin{equation}
\frac { \ZRA{\rho'}{\alpha'}(\tN) } { \ZRA{\rho}{\alpha}(\tN) } =
\frac
{
\sum_{\sS} \expc{\alpha' H(\sS)}
|\Pi_{\rho'}|
\fracd{\Pi_{\rho'}}{
|\Pi_{\rho'}| }
}
{
\sum_{\sS} \expc{\alpha H(\sS)}
|\Pi_{\rho}|
\fracd{\Pi_{\rho} }{
|\Pi_{\rho}| }
} .
\end{equation}
We finally divide both the numerator and the denominator, by
$ \sum \expb{\alpha H}| \Pi| $ to arrive at:
\begin{equation}
\frac { \ZRA{\rho'}{\alpha'}(\tN) } { \ZRA{\rho}{\alpha}(\tN) } =
\frac{\mean{\expc{(\alpha' - \alpha)H} \Pi_{\rho'}  /
|\Pi_\rho|}_{\alpha,|\rho|}}
{\mean{\Pi_\rho / \gle \Pi_\rho \gre}_{\alpha,|\rho|}} ,
\label{equ:ZRatioAbsRho}
\end{equation}
where $\mean{\dots}_{\alpha,|\rho|}$ refers to expectation values with respect
to
\begin{equation}
 \pi^{\{\alpha\}}_{|\rho|} (\sS | \tN) =
\expc{\alpha H(\sS)}|\Pi_\rho|,
\label{equ:BoltzmannAlphaAbs}
\end{equation}
which is an unnormalized probability distribution in the sample space $\Omega$
of old-site spin configurations of size $|\Omega| = 2^{221}$.
In the denominator of \eq{equ:ZRatioAbsRho}, we identify
\begin{equation}
\mean{\Pi_\rho / \gle \Pi_\rho \gre}_{\alpha, |\rho|} =
\mean{\text{sign}(\Pi_\rho)}_{\alpha, |\rho|},
\label{equ:Sign}
\end{equation}
in other words the average sign of the partition function.
In our application, this average sign usually differs markedly from zero, and
the relevant observable averages can still be computed.
The sign problem is well known from many low-temperature fermionic Monte Carlo
simulations. For $\alpha=0$, the partition function does not depend on the spins
$\sO$, and the average sign is given by
\begin{equation}
 \mean{\text{sign}\Pi_\rho}_{|\rho|}=
 \frac{
 \sum\limits_{\sS} \prod\limits_{k \in \NCAL} |(1 + \rho s_k)|\  \text{sign}\
\Pi_\rho
 }{
 \sum\limits_{\sS} \prod\limits_{k \in \NCAL} |(1 + \rho s_k)|
 } =
 \frac{
  \prod\limits_{k \in \NCAL} \sum\limits_{s_k}(1 + \rho s_k)
 }{
 \prod\limits_{k \in \NCAL} \sum\limits_{s_k}|(1 + \rho s_k)|
 } =
 \frac{2^{|\NCAL|}}{(2 \rho)^ {|\NCAL|}} =
\rho ^{-121},
\label{equ:AlphaZeroSign}
\end{equation}
a formula that will provides a useful check of our simulation results for small
values of $\alpha$. We will later discuss that a \emph{severe} sign problem
appears for some configurations $\tBN$ at large $\rho$, but they can be excluded
from our procedure.

\begin{algorithm}
    \newcommand{\algo}{alpha-rho-Metropolis}
    \captionsetup{margin=0pt,justification=raggedright}
    \begin{center}
        $\begin{array}{ll}
            & \PROCEDURE{\algo}\\
            & \INPUT{\sS, H, \Pi_\rho, \Pi_{\rho'}}\ \COMMENT{see
                                  \eq{equ:ZRatioAbsRho}}\\
            & \IS{i}{\sub{choice}(\SET{\sS})}\\
            & \IS{\Delta H}{-2  \sum_{M \in \MCAL: i \in I_M}
             K_{T_M} \prod_{j \in I_M} s_j}\\
            & \IS{\Gamma_\rho}{1};\ \IS{\Gamma_{\rho'}}{1} \\
            & \IF{i \in \NCAL}\\
            & \BRACE{
                  \IS{\Gamma_\rho}{(1 - \rho t_i s_i) / (1 + \rho t_i s_i) } \\
                  \IS{\Gamma_{\rho'}}{(1 - \rho' t_i s_i) / (1 + \rho' t_i s_i)}
            } \\
            & \IS{\Upsilon}{\ranb{0,1}}\\
            & \IF{\Upsilon < |\Gamma_\rho| \expb{\alpha \Delta H}}\\
            & \BRACE{
            \IS{s_i}{- s_i}\\
            \IS{H}{H + \Delta H}\\
            \IS{\Pi_\rho}{ \Gamma_\rho \Pi_\rho} \\
            \IS{\Pi_{\rho'}}{ \Gamma_{\rho'} \Pi_{\rho'}}
            } \\
            & \OUTPUT{\sS, H, \Pi_\rho, \Pi_{\rho'}} \\
            & \ENDPROCEDURE\
        \end{array}$
    \end{center}
    \caption{\sub{\algo}. One iteration of the single-spin-flip Metropolis
algorithm for sampling spin configurations $\sS$ with the distribution
$\pi^{\{\alpha \}}_{|\rho|}(\sS|\tN)$.
The algorithm provides for the computation of the ratio of
partition
functions in \eq{equ:ZRatioAbsRho} and of the average sign (see
\app{app:AccessComputer} for a Python implementation).}
\label{alg:\algo}
\end{algorithm}

\progg{alpha-rho-Metropolis} samples spin configurations $\sS$ from the
$(\alpha,\rho)$-dependent Boltzmann distribution of \eq{equ:BoltzmannAlphaAbs}.
It requires, initially (and optionally), the evaluation of $H$ through
\eq{equ:Hamiltonian}, but then updates $H$ through increments $\Delta H$. It
computes $\ZRA{\rho}{\alpha=1}(\tN)$ without systematic error. The ratio
computations of \eq{equ:ZRatioAbsRho} have a favorable propagation of the error
which, at $\alpha' = \alpha$ and $\rho' = \rho$, vanishes identically and then
grows as the two sets of parameters drift apart. The  downside of
\prog{alpha-rho-Metropolis} is the absence of correlation between
$\ZRA{\rho}{\alpha=1}(\tN)$ and $\ZRA{\rho}{\alpha=1}(\tN')$ for $\tN \neq
\tN'$, in other words the statistical independence of calculations of different
new $B$ spins (see \sect{sec:CorrelatedSampling} for possible correlated
samplings for $\tN \neq \tN'$).

\subsection{Correlated sampling}
\label{sec:CorrelatedSampling}

To locate the fixpoint to high precision, we must compute the change of  the
partition function $Z_\rho(\tBN | \tAN)$ and of other functions with small
variations of the interactions $K_T$, that is, of the hamiltonian $H$ to a
neighboring one $H'$. For the computation of gradients, the variations are
infinitesimal. Naively, under the rules of Gaussian error propagation, the error
bars in
$ Z_\rho^{\{H\}}$ and $ Z_\rho^{\{H'\}}$ add up when determining their
differences, so that the relative error explodes for $H \simeq H'$.
Correlated sampling overcomes this problem to a large part. For the computation
of the ratio of the two partition functions, it follows from an extension of
\eq{equ:ZRatioAbsRho} with the replacements
$ \alpha H \rightarrow H$ and $\alpha' H \rightarrow H'$ that:
\begin{equation}
\frac { \ZRA{\rho}{H'}(\tN) } { \ZRA{\rho}{H}(\tN) } =
\frac{\mean{\expb{H' - H } \Pi_{\rho}  /
|\Pi_\rho|}_{\alpha=1,|\rho|}}
{\mean{\Pi_\rho / \gle \Pi_\rho \gre}_{\alpha=1,|\rho|}} .
\label{equ:CorrelatedSampling}
\end{equation}
The quantities relating to $H'$ or to $H$ are now evaluated for identical
samples $\sS$. The statistical error vanishes if $H'$ and $H$ are the same, and
it remains manageable as long as $H'$ and $H$ have similar weights
for the same samples $\sS $. This eliminates the Gaussian error propagation
(see \sect{sec:TwoFactorFixpoint} for a practical application in our context).

Correlated sampling can be extended in many ways, most trivially to
small variations in $\rho$. We have not
tested the correlated sampling of partition
functions for two different configurations $\tBN$, that is, of the ratio
\begin{equation}
\frac { \ZRA{\rho}{H}(\tN') } { \ZRA{\rho}{H}(\tN) } =
\frac{\mean{\expb{H(\tN') - H(\tN) } \Pi_{\rho}  /
|\Pi_\rho|}_{\alpha, \tN, |\rho|}}
{\mean{\Pi_\rho / \gle \Pi_\rho \gre}_{\alpha, \tN, |\rho|}} .
\label{equ:CorrelatedSamplingtBN}
\end{equation}
This procedure would correlate the samples and greatly reduce the relative error
of the computation  for different values of $\tBN$, what
\eq{equ:CorrelatedSampling} does not achieve. We expect pairs of configurations
$\tBN$ which differ in one or two spins to be open to correlated sampling.
Propagation as in \sect{sec:CorrelatedSampling} would then allow to correlate
all partition functions $\ZRA{\rho}{H}$ with negligible statistical errors, and
therefore greatly improve the calculation of the \VR\ of
\sect{sec:ComputationalResults}. In this scheme, the integration over different
configurations $\tBN$ replaces, up to an irrelevant multiplicative constant, the
thermodynamic integration over the fictitious inverse temperature $\alpha$.

\section{Computational results}
\label{sec:ComputationalResults}
We have implemented \prog{alpha-rho-Metropolis}, as a proof of concept, in an
elementary Python script that is publicly accessible (see
\app{app:AccessComputer}). It reads in the $\SCAL$, $\OCAL$, and $\BCAL$
indices, as well as the factors. Following \REF{Wilson1975}, all new spins
$\tAN$ in the periphery are set to $+1$. We thus compute, using
\prog{alpha-rho-Metropolis}, the ratio
\begin{equation}
\Theta(\tBN | \tAN) =
\frac{
\ZRA{\rho}{\alpha=1} (\tBN | \tAN =
\SET{1 \TO 1})}
{\expc{H(\tBN | \tAN = \SET{1 \TO 1})}}
\stackrel{!}{=} \const,
\label{equ:Variation}
\end{equation}
which should be constant at a fixpoint. For our numerically rigorous integration
procedure, we do not expect an exact fixpoint to exist for a small number of
factor types. Nevertheless, a finite-$K$ fixpoint can be defined as the choice
of interactions $K_T$ minimizing the \quot{\VR}
\begin{equation}
\Psi(\SET{K_T}|\tAN) =
\frac{
\sqrt{
\mean{\Theta(\tBN | \tAN)^2} _{\tBN} -
      \glc\mean{\Theta(\tBN | \tAN}_{\tBN} \grc ^ 2}}
      {\sqrt{ \mean{\Theta(\tBN | \tAN}_{\tBN}  ^ 2}}
\label{equ:VariationRatio}
\end{equation}
The \VR $\Psi(\SET{K_T}|\tAN)$ is zero
when the fixpoint condition of
\eq{equ:Variation} is satisfied for the same constant for all inner spins
$\tBN$, in other words, at a fixpoint.
We can practically compute $\Psi$ by averaging over the $2^{11}$
configurations of $t$-spins in $\BCAL \cap \NCAL$, or a sufficiently large
subset thereof.

\subsection{Two factor types: integration path, average sign}
\proggg{alpha-rho-Metropolis} is tested for two factor types ($T=1$ and $T=2$ in
\fig{fig:FactorTypes}) with non-vanishing interactions $K_1$ and $K_2$ that, for
coherence with Wilson's notations, we call $K$ and $L$. We then deal with $800$
factors for the old spins $s_i: i \in \SCAL$ and $420$ factors for the new ones.
For this two-factor-type model, Wilson has obtained the non-trivial fixpoint in
second-order perturbation theory as
\begin{align}
K^* &= (2\rho^2 + \rho^4)^{-1} \label{equ:SecondOrderK} \\
L^* &= \rho^2 (2\rho^2 + \rho^4)^{-2} \label{equ:SecondOrderL}
\end{align}
(see \app{app:WilsonVITwoParameter} for a self-contained derivation).
For a coupling parameter $\rho=1.04$, the perturbation-theory fixpoint is at
$K^*=0.3, L^*=0.10736$. For these interactions, we have studied in
detail the average \quot{fermion} sign of \eq{equ:Sign} as a function of
$\alpha$ and $\rho$. For $\rho$ somewhat larger than $1.006$, the average
sign approaches zero and the sign problem thus becomes severe for $\alpha \to
0$ (see \eq{equ:AlphaZeroSign}). The integration path (the chain of pairs
$(\alpha, \rho)$ and
$(\alpha', \rho')$ leading towards $\alpha=0$)
thus has to
move to smaller $\rho$, as shown in \subfig{fig:AverageSign}{a} for a specific
choice of $\tBN$ spins. For locating the fixpoint, we need to compute
the partition function for a number of different $\tBN$ spins. We have
monitored the sign problem at $\alpha=1$ for the $2^{11}$ different $\tBN$
spin configurations and found it to be manageable for almost all choices at
$\rho=1.05$, and for all choices of $\tBN$ for smaller $\rho$ (see
\subfig{fig:AverageSign}{b}). A Python script for this computation is available
in \app{app:AccessComputer}.

\draftfigure[14cm]{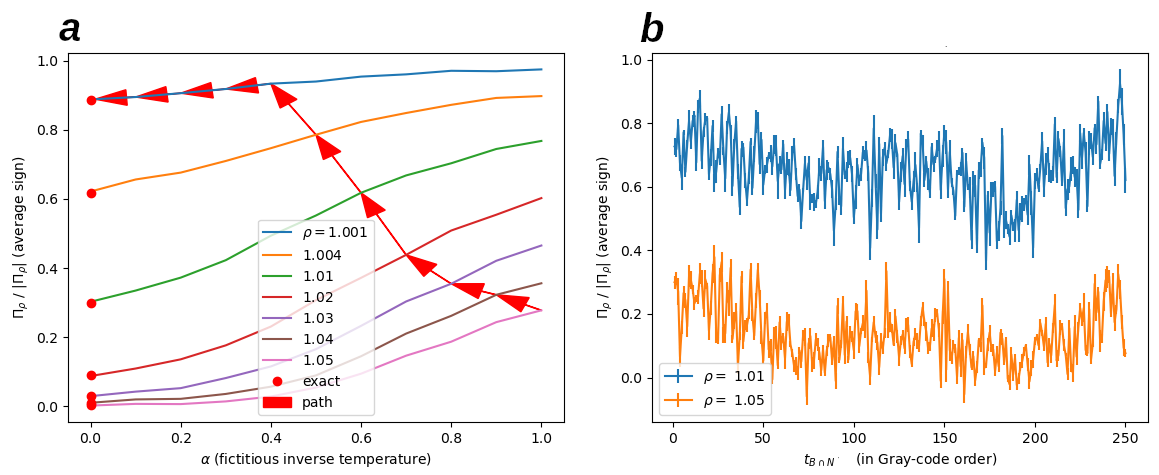}{Average sign of
\prog{alpha-rho-Metropolis} for the two-factor-type model with interactions
$(K, L) = (0.3,0.10736)$. \subcap{a} Average sign and
thermodynamic integration path ($\tBN=2$, $\tAN = \SET{1\TO 1}$)
connecting
$(\alpha,\rho)=(1,1.05)$ to the known partition function at
$(\alpha, \rho) = (0, 1.001)$, with exact values (red dots)  from
\eq{equ:AlphaZeroSign}.
\subcap{b} Average sign at $\alpha=1$ for the
spin configurations $\tBN = \SET{1\TO 250}$, for two different coupling
parameters $\rho$.
}{fig:AverageSign}

\subsection{Two factor types: correlated sampling, fixpoint}
\label{sec:TwoFactorFixpoint}

To illustrate the force of correlated sampling, we consider,  in the
two-factor-type model, the calculation of partition functions
$\ZRA{\rho}{\alpha=1}$ for the interactions $(K,L)= (0.301,0.10736)$ and
the nearby $(K,L)= (0.3,0.10736)$. Independent computations with $\sim
\fpn{2}{8}$ samples distributed over the thermodynamic integration from
$\alpha=1$ to $\alpha=0$ with $10$ intermediate values of $\rho$ and $\alpha$
yield an estimate of the ratio of these partition functions as $1.388 \pm 0.04$.
Correlated sampling (using \eq{equ:CorrelatedSampling}) requires no
thermodynamic integration, and $\fpn{1}{7}$ samples
yield the ratio of these partition functions to $1.3994 \pm 0.0001$. It thus
gains several orders of magnitude in precision (see \app{app:AccessComputer}
for example programs retracing this computation).

For our proof-of-concept computation, we have not implemented our
correlated-sampling ideas but simply computed the \VR $\Psi\glc(K, L)|\tBN =
\SET{1 \TO 1}\grc$
from independently computed partition functions (using thermodynamic
integration over $10$ intermediate values of  $\rho$ and $\alpha$)
for a grid of values of $K$ and $L$.
In \fig{fig:WilsonVIRough}, we plot the \VR $\Psi$ of \eq{equ:VariationRatio}
for $\rho=1.04$. The
\VR is smallest for $(K,L) \simeq
(0.31,0.11)$, in good agreement with perturbation theory. To simplify the
calculation, and given the small number of interactions $K_T$ to be fitted,
$\Psi$ was computed from six of the $\tBN$ configurations encoded in
\tab{tab:GrayCodeTBN} rather than for all $2048$.

\draftfigure[7cm]{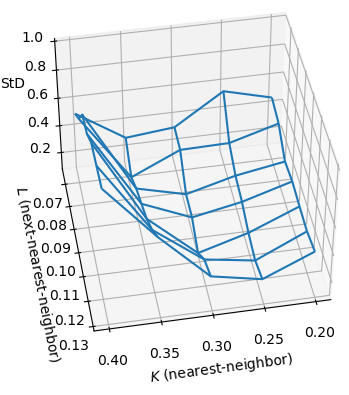}{Plot of the \VR
$\Psi(K,L|\SET{1\TO 1})$ for the two-factor-field model on a grid of
interactions $(K, L)$ for correlation parameter $\rho=1.04$.
The variations are smallest for $(K, L) \simeq (0.31,
0.11)$, in good agreement with second-order perturbation theory.
}{fig:WilsonVIRough}

\subsection{Fourteen factor types}
\label{sec:FourteenFactorTypes}
For the fourteen factor types in \fig{fig:FactorTypes}, we have $9447$ factors
for the old spins and $4795$ factors for the new spins. For this problem, we
have restricted ourselves to the computation of the average sign for the $14$
values of $K_T$ given in \cite[Table III]{Wilson1975} (see
\fig{fig:AverageSign14}), which appears roughly as for the two-factor-field
model in  \fig{fig:AverageSign}. Globally, we conclude that the sign problem
remains under control.

\draftfigure[8cm]{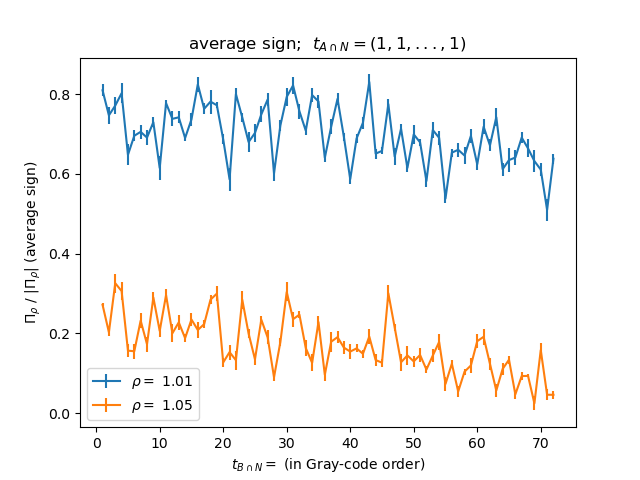}{Average sign of
\prog{alpha-rho-Metropolis} for $\tBN \in \SET{1 \TO 72}$ (see
\app{app:Details} for the enumeration), in the fourteen-factor-type model with
interactions $K_T$ as in \cite[Table III]{Wilson1975}
}{fig:AverageSign14}

\section{Conclusions}
In this work we have retraced, on the level of a proof of concept, the
classic real-space-renormalization computation by Wilson~\cite{Wilson1975},
using state-of-the-art Monte Carlo techniques~\cite{SMAC}. Our computations
can be generalized in many ways, and our publicly accessible Python
scripts (see \app{app:AccessComputer}) can be extended in directions that we
have indicated.  They can be rendered many orders of magnitude more
precise, through better codes, more adapted computer languages, but mostly
through the systematic use of correlated sampling.

Kadanoff and Wilson, half a century ago, modified the original decimation
procedure at the core of real-space renormalization in order to allow for
non-trivial algebraic scaling of spin--spin correlation functions. In our
computational framework, this brings in aspects of fermionic Monte Carlo and of
its notorious sign problem~\cite{Blankenbecler1981} which, as we have shown
here, can be handled with relative ease. It is unclear to us whether the
modified decimation procedure is itself free of contradictions, and if so,
whether it is unique and can be generalized beyond the two-dimensional Ising
universality class. Our numerical routine for computing the fixpoint uses key
Monte Carlo techniques (thermodynamic integration, fermion-sign problem,
correlated sampling). Much
understanding will have to be built up, though. We have noticed, for example,
that the sign problem is more severe for spins $\tAN$ differing from Wilson's
choice $\tAN=\SET{1 \TO 1}$. The significance of the effectively fermionic
partition function within classical statistical mechanics needs to be better
understood, just as the very definition of a fixpoint as a minimal-variance
point for a finite number of factor types which we expect to converge towards a
\emph{bona fide} fixpoint in the limit of infinite factor types.

\section*{Acknowledgements}
We are grateful to Slava Rychkov for support and helpful discussions.

\paragraph{Funding information}
W. K. acknowledges generous support by the Leverhulme Trust.
\appendix
\numberwithin{equation}{section}

\section{Perturbative fixpoint for two factor types}
\label{app:WilsonVITwoParameter}
In this appendix, we derive Wilson's fixpoint equations in second-order
perturbation theory for the two-factor-type model. We consider a
sufficiently large
lattice to accomodate all hopping terms to second order (see
\subfig{fig:KadanoffTwo}{a}).

\draftfigure[10cm]{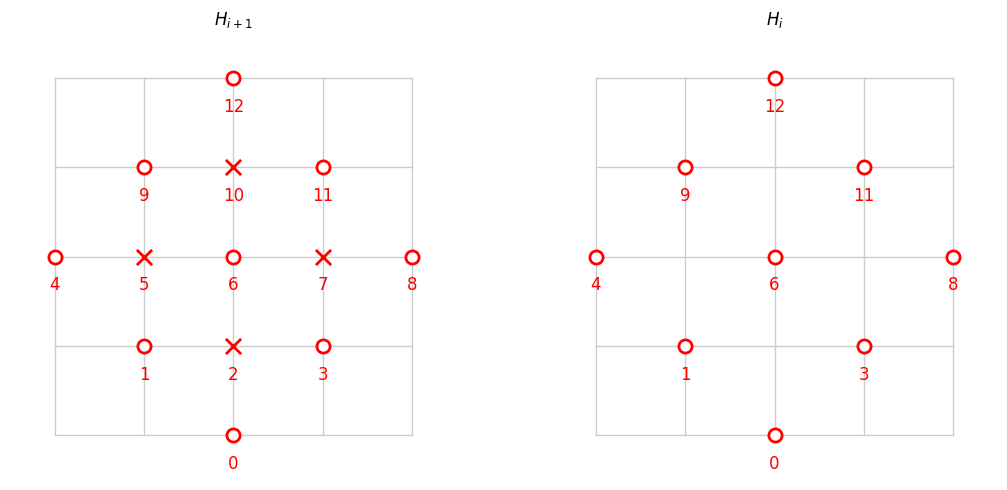}{Kadanoff decimation for the two-dimensional
Ising model with $K$ and $L$ interactions on a small lattice.}{fig:KadanoffTwo}

In Kadanoff's orginal decimation (see \sect{sec:OriginalDecimation}),
on this lattice, we integrate
out the \quot{only old} spins $s_2, s_5, s_7, s_{10}$,  in order to arrive at
the renormalized $H_{i+1}$ on the lattice of new spins
\subfig{fig:KadanoffTwo}{b}.
We have
\begin{equation}
\exp{H_{i+1}} =
\!\!\!\!\!\!\!\!\!
 \sum_{s_2= \pm 1,s_5, s_7, s_{10} = \pm 1}
\!\!\!\!\!\!\!\!\!
 \expb{K_i s_1 s_2 + K_i s_2 s_3 \PLUSPLUS L_i s_0 s_1 + L_i s_2 s_7
\PLUSPLUS L_i s_{11} s_{12}}.
\label{equ:SecondOrderExpansionA}
\end{equation}
In \eq{equ:SecondOrderExpansionA}, terms in $K_i$ have one old spin and one new
spin, while the terms in $L_i$ have either two old spins or two new spins.
We expand the exponential to second order, and keep only terms quadratic in
$K_i$ or else linear in $L_i$. This gives
\begin{multline}
\expb{H_{i+1}} =
 \sum_{s_2,s_5, s_7, s_{10}}
 \left[
 1 +
 \glb K_i s_1 s_2 + K_i s_2 s_3  \PLUSPLUS L_i s_0 s_1 + L_i s_2 s_7 \PLUSPLUS
L_i s_{11}s_{12} \grb
 \right. \\
 \left.
 + \half
 \glb K_i s_1 s_2 + K_i s_2 s_3 \PLUSPLUS
\grb^2
 \right].
\label{equ:SecondOrderExpansionB}
\end{multline}
The constant term $1$ sums up to $2^4 = 16$. The linear terms in $K_i$ sum
to zero. The linear terms in $L_i$ sum zu zero if they connect old spins, and
they sum up to $16$ if they connect new spins. Non-vanishing quadratic terms
are  of two types. Those connecting a chain of spins are as
\begin{equation}
 \half K_i^2 \glb s_6 s_7 s_7 s_{11} + s_6 s_{10}s_{10} s_{11} \grb,
\end{equation}
but each term $s_6 s_7 s_7 s_{11}$ appears twice, once with the $s_6 s_7$ in
the first of the two braces $(\dots)(\dots)$, and once in the second. The
terms coupling $s_6$ to $s_{11}$ sum up to $16\times \half \times 2 \times
2 = 16 \times 2$.
Quadratic terms, such as $\half K_i^2 (s_0 s_2) ^2$, that go from one spin and
back, sum up to $ 16 \times \half \times  16 K_i^2$, as there are $16$ edges
on this lattice, and in addition $16$ spin configurations to be summed over.
This yields:
\begin{equation}
\expb{H_{i+1}} = 16 \glc 1 +  \cdots \glb 2 K_i^2 + L_i \grb s_6 s_{11}
\PLUSPLUS K_i^2 s_6 s_{12} \PLUSPLUS 8 K_i^2 \grc.
\end{equation}
After re-exponentiation, the $s_6,s_{11}$ interaction corresponds to $K_{i+1}$
and the $s_6 s_{12}$ interaction to $L_{i+1}$, reproducing Wilsons eqs. (VI.18)
and (VI.19). The results would be identical for any larger lattice, as we made
at most two hops.

In the modified decimation (see \sect{sec:ModifiedDecimation}),
all spins $\SET{s_0 \to s_{12}}$ are summed
over in  $\expb{H_i}$, and the coupling with the spins $t$ of the hamiltonian
$H_{i+1}$ is expressed through $\rho$. This gives
\begin{equation}
 \expb{H_{i+1}} = \sum_{s_0 \TO s_{12}} \expb{K_i s_0 s_2 \PLUSPLUS + L_i s_2
s_7 \PLUSPLUS} \prod_{k=0,1,3,4,6,8,9,11,12} \frac{1 + \rho t_k s_k}{2}.
\label{equ:ModifiedKadanoff1}
\end{equation}
We expand the exponential to second order. The constant term gives $2^{13}/2^9
= 2^4$, because in the product term of \eq{equ:ModifiedKadanoff1}, only the
$\half$ terms survive. To linear order, all the $K_i$ terms disappear,
because of the presence of an isolated old spin. There are linear $L$ terms,
such as
\begin{equation}
 L s_6 s_{11} (1 + \rho t_6 s_6) (1 + \rho t_{11} s_{11} = L t_6 t_{11} \rho^2,
\end{equation}
again multiplied by a factor $16$.
To second order from $6$ to $11$, we have a term
\begin{equation}
 \sum_{s_0 \TO s_{12}} \frac{1}{2^9} \half K_i^2
 \glb
 s_6 s_7 s_7 s_{11} + s_7 s_{11} s_6 s_7
 s_6 s_{10} s_{10} s_{11} + s_{10} s_{11} s_6 s_{10}
 \grb s_6 s_{11} \rho^2 t_6 t_{11} = 16 \times 2 K_i^2 \rho^2 t_6 t_{11},
\end{equation}
whereas the term coupling $t_6 $ and $t_{12} $ is half this one.
This yields:
\begin{equation}
\expb{H_{i+1}} = 16 \glc 1 +  \cdots \glb 2 K_i^2 + L_i \grb \rho^2 t_6 t_{11}
\PLUSPLUS K_i^2  \rho^2 t_6 t_{12} \PLUSPLUS  \grc.
\end{equation}
After re-exponentiation,
the $t_6,t_{11}$ interaction corresponds to $K_{i+1}$ and
the $t_6 t_{12}$ interaction to $L_{i+1}$, reproducing
Wilsons eqs. (VI.45) and (VI.46), which correspond to our
\eqtwo{equ:SecondOrderK}{equ:SecondOrderL}.

\section{Enumeration of new central spins}
\label{app:Details}

Configurations of the $11$ new spins $\tBN$ in the central regions are
required for the study of the sign problem (the computation of the average
sign) and for the minimization procedure. For concreteness, we
enumerate them  using the Gray code (see \tab{tab:GrayCodeTBN}).
The corresponding Python script is available in our open-source
repository (see \app{app:AccessComputer}).
\begin{table}[h]
    \centering
		\begin{tabular}{rc}
		 number & $t_{89}, t_{90}, \dots, t_{133}, t_{134}$ \\
			\hline
1 & $-----------$ \\
2 & $----------+$ \\
3 & $---------++$ \\
4 & $---------+-$ \\
.. & \dots      \\
820 & $-+-+-+-+-+-$ \\
.. & \dots      \\
1366&  $+++++++++++$ \\
.. & \dots      \\
1639 & $+-+-+-+-+-+$ \\
.. & \dots      \\
2048 & $+----------$
   \end{tabular}
    \caption{Numbering scheme for the new spins $\tBN$ in the central region,
following the Gray code. The first column of this table corresponds to the
$x$-axis in \subfig{fig:AverageSign}{b} (see \app{app:AccessComputer} for a
Python script). }
\label{tab:GrayCodeTBN}
\end{table}

\section{Access to computer programs}
\label{app:AccessComputer}
This work is accompanied by the \texttt{MCRealSpace} software package, which is
published as an open-source project under the GNU GPLv3 license.
\texttt{MCRealSpace} is available on GitHub as a part of the JeLLyFysh
organization. The package contains Python scripts for the algorithms discussed
in this work, as well as the data files encoding the Wilson patch, including
the factors. The url of the repository is
\url{https://github.com/jellyfysh/MCRealSpace.git}.

\end{document}